\documentclass[epj]{svjour}
\usepackage{graphics}
\newcommand{\pcm}{\ensuremath{p_\mathrm{cm}}}
\newcommand{\h}{\ensuremath{\eta}}
\newcommand{\dd}{\mbox{\rm d}}

\def\fmn#1#2{\mbox{${\textstyle \frac{#1}{#2}}$}}

\renewcommand{\deg}{\ensuremath{^{\circ}}}
\newcommand{\hef}{\mbox{\ensuremath{^4\mathrm{He}}}}
\newcommand{\het}{\mbox{\ensuremath{^3\mathrm{He}}}}
\newcommand{\half}{\mbox{${\textstyle \frac{1}{2}}$}}           % 1/2
\newcommand{\bmath}[1]{\mbox{\boldmath ${#1}$}}

\newcommand{\hefx}{\mbox{${dd}\,\to\,^4\mathrm{He}\,{X}$}}
\newcommand{\hetn}{\mbox{${dd}\,\to\,^3\mathrm{He}\,{n}$}}
\newcommand{\heteta}{\mbox{${pd}\,\to\,^3\mathrm{He}\,\eta$}}

\newcommand{\hefeta}{\mbox{${dd}\,\to\,^4\mathrm{He}\,\eta$}}

\begin{document}
\title{Near threshold $\eta$ meson production in the \hefeta~reaction}
\author{A.~Wro\'nska\inst{1,2,}\thanks{\email{A.Wronska@fz-juelich.de}}
\and V.~Hejny\inst{1}%
\and C.~Wilkin\inst{3}%
\and S.~Dymov\inst{1,4}%
\and C.~Hanhart\inst{1}%
\and V.~Komarov\inst{4}%
\and H.R.~Koch\inst{1}%
\and A.~Kulikov\inst{4}%
\and A.~Magiera\inst{2}%
\and T.~Mersmann\inst{5}%
\and S.~Mikirtytchiants\inst{6}%
\and H.~Ohm\inst{1}%
\and D.~Prasuhn\inst{1}%
\and R.~Schleichert\inst{1}%
\and H.J.~Stein\inst{1}%
\and H.~Str\"oher\inst{1}%
}                     % Do not remove
%
%\offprints{A.~Wronska}          % Insert a name or remove this line
%
\institute{
  Institut f\"ur Kernphysik, Forschungszentrum J\"ulich, 52425~J\"ulich, Germany
  \and
  Institute of Physics, Jagiellonian University, Reymonta~4, 30--059~Cracow, Poland
  \and
  Physics Department, University College London, London WC1E 6BT, U.K.
  \and
  Laboratory of Nuclear Problems, Joint Institute for Nuclear Research, 141980 Dubna, Russia
  \and
Institut f\"ur Kernphysik, Universit\"at M\"unster, 48149 M\"unster, Germany
  \and
  High Energy Physics Department, PNPI, 188350 Gatchina, Russia
}
\date{Received: / Revised version: date}
\abstract{The \hefeta{} reaction has been investigated near
threshold using the ANKE facility at COSY--J\"ulich. Both total
and differential cross sections have been measured at two excess
energies, $Q=2.6$~MeV and 7.7~MeV, with a subthreshold measurement
being undertaken at $Q=-2.6$~MeV to study the physical background.
While consistent with isotropy at the lower energy, the angular
distribution reveals a pronounced anisotropy at the higher one,
indicating the presence of higher partial waves. Options for the
decomposition into partial amplitudes and their consequences for
determination of the $s$-wave $\eta$-$\alpha$ scattering length
are discussed.
 % end of PACS codes
} %end of abstract
\maketitle
\section{Introduction}
\setcounter{equation}{0}
\label{sec:intro}%
Interest in the physics of the \h{}~meson underwent a revival in
the 1980s when it was discovered that the \h--nucleus interaction
is so strong and attractive that the existence of
\h--nucleus~quasi--bound states was
hypothesised~\cite{Liu86,Haider86,Chiang91}. The original
predictions of such states concerned heavier nuclei ($A>11$) but
direct searches for them in the cases of lithium, carbon, oxygen
and aluminium proved inconclusive~\mbox{\cite{Chrien88,Lieb00}}.
More recent theoretical approaches, using more attractive
\h--nucleus interactions, gave positive predictions for much
lighter nuclei, \emph{e.g.}\ the helium
isotopes~\mbox{\cite{Wilkin93,Wycech95,Fix02,Rakityansky95,Rakityansky96}}.
Data on the \heteta~and \hefeta~reactions obtained at the SATURNE
accelerator~\cite{Berger88,Mayer96,Frascaria94,Willis97} and other
laboratories~\cite{Betigeri00,Bilger02,Adam05} have been
interpreted by some authors as suggesting that the \h--$^4$He, and
perhaps even \h--$^3$He, systems could support bound
states~\cite{Willis97,Tryasuchev97}. Attempts have also been made
to photoproduce the latter~\cite{Pfeiffer}, though the analysis is
not completely unambiguous~\cite{Hanhart:2004qs}.

It is important to stress that \h--nucleus bound states should
first occur in the $s$-wave. Thus, any calculation exploiting
information extracted just from total cross sections relies on the
implicit assumption of purely $s$-wave production in the proximity
of the reaction threshold. Verification of this assumption
requires the measurement of angular distributions and polarisation
observables as well as total cross sections. Differential cross
sections do exist for \heteta{} close to
threshold~\mbox{\cite{Mayer96,Betigeri00,Bilger02,Adam05}}, and
these indicate the onset of higher partial waves at
$Q\approx15$~MeV. For the \hefeta{} process, data on the total
cross section are restricted to small excess energies ($Q<9$~MeV)
and no data whatsoever are available on differential cross
sections.

\section{Experiment}%
\setcounter{equation}{0}
\label{sec:exper}%
The experiment~\cite{Wronska05}, aiming at the determination of
total cross sections and angular distributions for \hefeta{} close
to threshold, was performed in the Institut f\"ur Kernphysik of
the Forschungszentrum J\"ulich. The measurement was carried out at
the ANKE facility~\cite{Barsov01}, located at an internal target
position of the COSY synchrotron, for three beam momenta,
2.328~GeV/c, 2.343~GeV/c and 2.358~GeV/c. The first of these lies
below the reaction threshold and was undertaken to study the shape
of the physics background. The other two correspond to excess
energies of 2.6~MeV and 7.7~MeV, respectively\footnote{The
translation between excess energies and the corresponding beam
momenta has been done assuming an \h{} mass of
$m_{\eta}=547.3$~MeV/c$^2$~\cite{gem05}.}. With around $5\times
10^{10}$ deuterons circulating in the COSY ring and a D$_2$
cluster jet target~\cite{target}, a luminosity of about
\mbox{$4\times10^{30}$~s$^{-1}$cm$^{-2}$} was achieved.

\begin{figure}
\resizebox{0.5\textwidth}{!}{
  \includegraphics{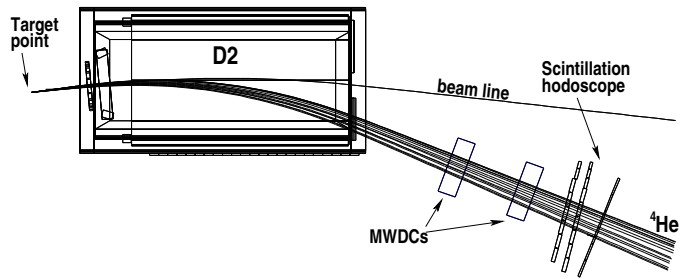}  % created from withsw.fig
} \caption{Layout of the forward detection system of ANKE used in
the experiment. Several trajectories of $\alpha$-particles
originating from the \hefeta{} reaction are depicted.}
\label{fig:withsw}
\end{figure}

The measurement assumed detection of~\hef{} particles and
reconstruction of their momenta, followed by a missing--mass
analysis of the remaining reaction products. The detection system
was based on three magnets D1-D3 forming a chicane in the COSY
ring (full setup is presented {\it e.g.}~in fig.2 of~\cite{Barsov01}),
with the target position located between D1 and the
spectrometer magnet D2. In the layout used in the experiment (see
fig.~\ref{fig:withsw}), only the forward detection system located
between the D2 and~D3 magnets was exploited. This comprised two
multiwire drift chambers (MWDC) for track reconstruction and three
layers of the scintillation hodoscope allowing time--of--flight
and energy--loss determinations.

Simulations showed that the ANKE~acceptance is nearly 100\% up to
$Q\approx 6.6$~MeV and full coverage in polar angle is assured up
to $Q\approx 90$~MeV. The angular resolution (in cms) for the
$\alpha$-particles stemming from \hefeta{} was determined to be
not worse than 19\deg{} for the lower energy ($Q_1$) and 11\deg{}
for the higher ($Q_2$). Moreover, the resolution in the transverse
momentum component was about four times better than that of the
longitudinal component. Thus, investigation of the transverse
momentum spectrum provides direct information on the excess energy.

The application of a dedicated energy--loss--based trigger was
necessary in order to obtain satisfactory background suppression
for the data acquisition system. Since all scintillation counters
were read out from both sides, this required special integrating
modules, allowing summation and integration of two analogue
signals. The discrimination threshold was set below the
\mbox{$\Delta E$--$p$} band of the \het. Additionally, a sample of
data was collected with a minimum bias trigger.

\begin{figure}
\resizebox{0.5\textwidth}{!}{
  \includegraphics{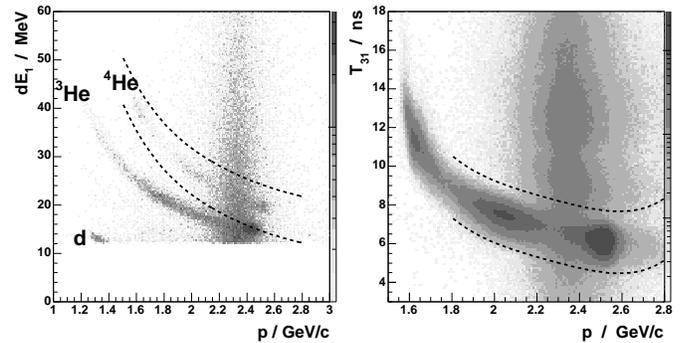}  % created with HeCuts.C
} \caption{Typical shapes of the $\Delta E_i$--$p$ (left) and
$T_{j1}$--$p$ (right) cuts applied in the selection of the \hef{}
particles.} \label{fig:cuts}
\end{figure}

\begin{figure}
\centering
\resizebox{0.5\textwidth}{!}{
  \includegraphics{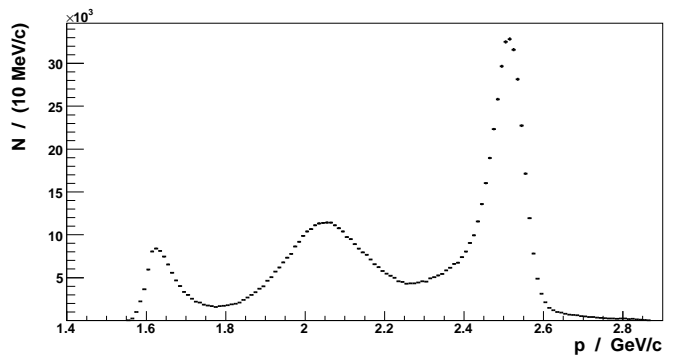}
} \caption{Inclusive momentum distribution of \hef{} particles
  measured at excess energy $Q_1$.} \label{fig:momentum}
\end{figure}

There was strong contamination in the raw data from protons
resulting from deuteron break-up since these have rigidities very
close to those of the \hef{} particles of interest. In order to
ensure a clean selection of \hef{} events, it was necessary to
apply cuts on all $\Delta E_i$--$p$~spectra ($i=1,2,3$) and
$T_{j1}$--$p$ ($j=2,3)$ with the reference time being given by the
signal from the first hodoscope layer. The cut shapes were
determined by dividing the three-dimensional spectra into slices
and fitting the contents with Gaussians. The cut widths were taken
to be at least three standard deviations and, in case of
ambiguities (in the region of the break-up background), the width
was interpolated between neighbouring regions with lower
background content. Typical shapes of the cuts applied are shown
in fig.~\ref{fig:cuts}.
The overall suppression factor from minimum bias trigger to
software \hef{} selection amounted to about $10^5$. The final
momentum spectrum depicted in fig.~\ref{fig:momentum} exhibits a
three--peak structure, similar to those reported in
ref.~\cite{Banaigs76}. This is a reflection of the well--known
ABC~effect~\cite{ABC}, which leads to kinematic enhancements of
the two--pion mass spectrum at both small masses (the outer peaks)
and at maximum mass (the central peak)~\cite{GFW}.
\begin{figure}
\resizebox{0.5\textwidth}{!}{
  \includegraphics{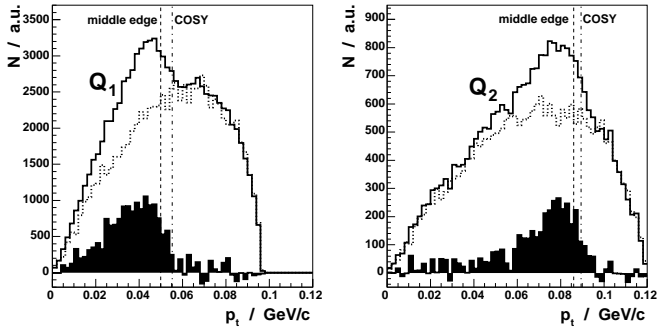}
} \caption{Distributions of transverse momenta of identified
$\alpha$--particles obtained for nominal beam momenta of
2.345~GeV/c (left, solid line) and 2.360~GeV/c (right, solid line)
for events with $m_{X}>0.54$~GeV/c$^2$. On top of them the
phase-space-scaled subthreshold data are drawn with dotted lines.
The difference, corresponding to events just with \h~production,
is represented by filled histograms. Vertical lines indicate
\pcm~calculated from the nominal beam momenta and that deduced
from the presented spectra.} \label{fig:pt}
\end{figure}

The solid lines in fig.~\ref{fig:pt} represent data obtained above
threshold at $Q_1$ and $Q_2$ while the dotted lines indicate
background. For both of the energies studied the background
measured at $Q_0$ was scaled according to the available phase
space and luminosity ratio~\cite{Hibou99}, the latter being
determined from the ratio of the histogram sums outside of the
\h{} peak region. The filled histograms result from subtraction of
the background spectra and should thus represent pure
\h{}~signals.

\begin{figure*}[htb]
\resizebox{\textwidth}{!}{
  \includegraphics{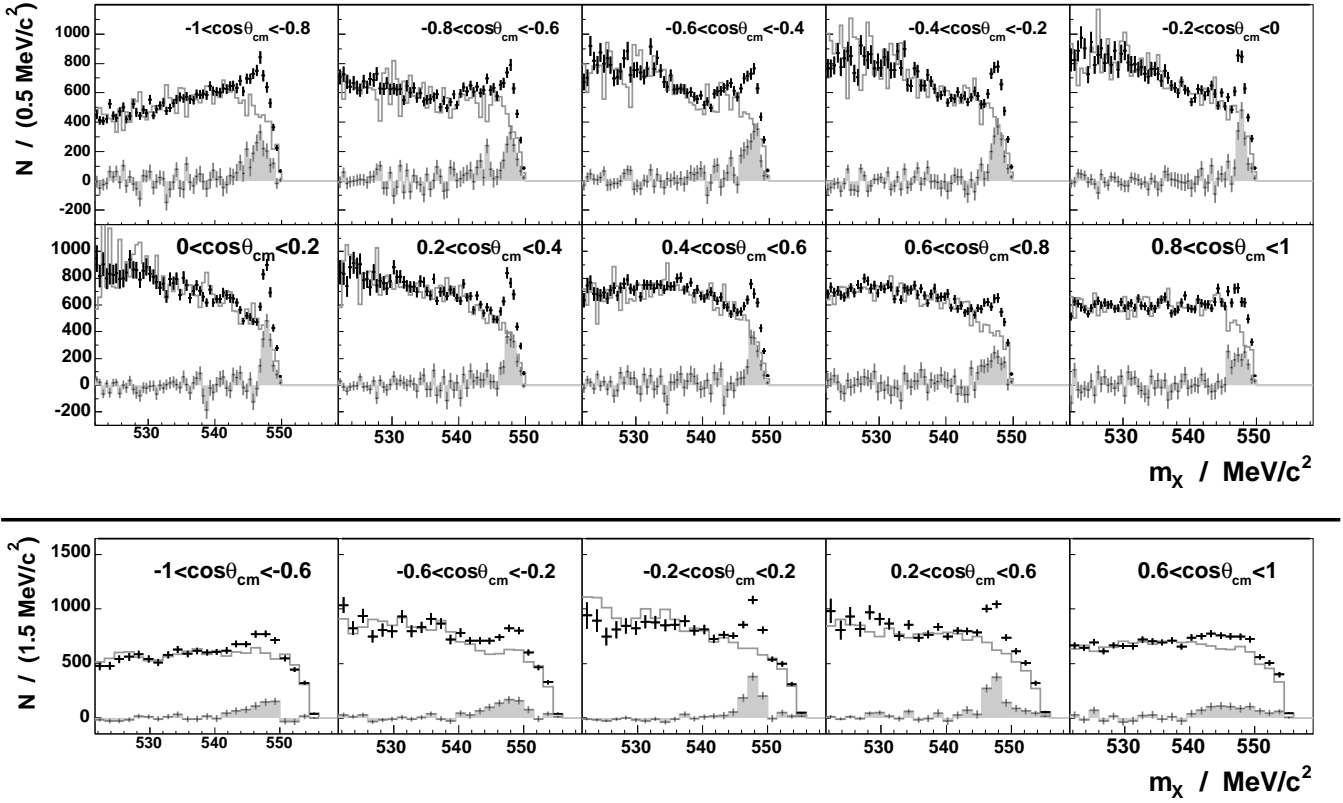} %macro subtract.C
} \caption{Missing-mass distributions for the \hefx{} reaction in
bins of cms angle at excess energies $Q_1$ (upper panels) and
$Q_2$ (lower panels). The scaled background, shown by the grey
lines, is subtracted from the data to leave the desired \h{}
peaks. } \label{fig:mm_bins}
\end{figure*}

Analysis of the transverse momentum spectra shown in
fig.~\ref{fig:pt} provides information on the cms momentum of the
outgoing particles, and thus the excess energy which does not rely
on knowing the mass of the $\eta$--meson precisely. The
simulations show that the final cms momentum can be determined
from the middle of the steeper edge of the transverse momentum
distributions. In the left panel of fig.~\ref{fig:pt}, the middle
of the edge occurs at 50~MeV/c rather than 55~MeV/c, as calculated
from the nominal beam momentum using the GEM value of
$m_{\eta}$~\cite{gem05}. This difference, which depends upon the
assumed value of $m_{\eta}$, indicates that the real beam momentum
in the $Q_1$ case was about 2~MeV/c lower than the nominal value,
which is within the precision of the absolute COSY settings. The
relative beam momenta are, however, known with an order of
magnitude better precision so that the shift found at $Q_1$ was
assumed to be valid also at the other two energies. The spectra on
the right of~fig.~\ref{fig:pt} do not contradict this hypothesis.

The deduced values of the cms momenta and the corresponding excess
energies are:
\begin{eqnarray*}
  &&Q_0 = (-2.6 \pm 0.6) \textrm{ MeV ,}\\
  p_\mathrm{cm,1} &= (50\pm 5) \textrm{ MeV/c,~~~}&Q_1
  = \phantom{-}(2.6 \pm 0.6) \textrm{ MeV,}\\
  p_\mathrm{cm,2} &= (86\pm 4) \textrm{ MeV/c,~~~}&Q_2
  = \phantom{-}(7.7 \pm 0.8) \textrm{ MeV.}
\end{eqnarray*}

Acceptance and resolution corrections were determined through
GEANT simulations~\cite{geant3}, taking into account the influence
of physics processes (small angle scattering, energy loss,
\emph{etc.}) as well as the setup features (geometry, extended
target, finite position resolution of the MWDCs, \emph{etc.}). The
acceptance $A(m_{X},\cos\theta_\mathrm{cm})$ was expressed in terms
of the missing mass and the cosine of the cms polar angle. As
starting values, we used data on inclusive \hef~production in $dd$
collisions~\cite{Banaigs76} but, to ensure good statistics also in
the cells that were poorly populated in the event generator, this
was supplemented with a sample of uniformly distributed events.
For the total event distribution this introduces only a small
change, but it helps significantly in the reduction of the final
statistical uncertainty. For the above-threshold energies,
separate samples of isotropically produced \h{} events were
generated, but these were only used to correct the angular
distributions.

To find the angular distributions of the \hef{} from the \hefeta{}
reaction, events were divided into angular bins and missing mass
spectra were considered separately for each bin, the scaled
background events being treated in the same way. For the lower
energy above threshold, the statistics allowed a division into ten
angular bins of equal width in $\cos\theta_\mathrm{cm}$. For the
higher energy, where the statistics and missing mass resolution
were several times worse, the data were divided into five
intervals. In the results, shown in fig.~\ref{fig:mm_bins}, the
background is corrected using the \hefx{} simulation. The
resulting subtracted spectra should represent pure \h~signals
(filled histograms in fig.~\ref{fig:mm_bins}) though influenced by
acceptance and resolution. The effects of these were unfolded
using information from the simulation of \hefeta~events.

The total luminosities were estimated by comparing data with the
inclusive differential \hefx{} results of ref.~\cite{Banaigs76},
which have an absolute uncertainty of about 15\%. The central bump
in the three--peak structure of the \hef{} momentum distribution
was parametrised by a Gaussian with parameters depending on the
polar angle and beam momentum. The integrated luminosities for the
three excess energies were:
\begin{eqnarray*}
L_0 & = &\phantom{1}(316 \pm 3_\mathrm{\,stat})\,\textrm{nb}^{-1}\:,\\
L_1 & = &(1566\pm 21_\mathrm{\,stat})\,\textrm{nb}^{-1}\:,\\
L_2 & = &\phantom{1}(299\pm
5_\mathrm{\,stat})\,\textrm{nb}^{-1}\:. \label{eq:lumiHe4}
\end{eqnarray*}

As checks on the luminosity, values were also determined from
yields of the \hetn{} reaction compared with the published data of
ref.~\cite{Bizard80}, the event selection and acceptance
correction being performed in a way similar to that described for
\hef{}. The numbers obtained from this analysis (317$\,$nb$^{-1}$,
1425$\,$nb$^{-1}$ and 255$\,$nb$^{-1}$ at the three energies) are
consistent with those determined from the \hefx{} data within the
overall uncertainties. However, the latter reaction is preferable
for our purposes since it involves the detection of \hef~particle
with momenta similar to those arising from \hefeta. By normalising
to the \hefx~process, uncertainties originating from the detection
setup efficiency cancel completely and those in the acceptance at
least partially.

\section{Results}
\label{sec:resul} \setcounter{equation}{0}

The analysis of the experimental data presented in the previous
section yielded the following total cross sections
\begin{eqnarray}
\sigma_1 & =& \left(13.1\pm 0.7_\mathrm{\,stat}\pm
1.8_\mathrm{\,syst}\right)\:\textrm{nb}\:,\nonumber\\
\sigma_2 & =& \left(16.4\pm 1.0_\mathrm{\,stat}\pm
2.1_\mathrm{\,syst}\right)\,\textrm{nb}\:.
\end{eqnarray}
The systematic errors quoted here have been estimated by varying
the conditions of the analysis within their uncertainties and do
not include the 15\% uncertainty in the luminosity, which does not
affect the relative size of the cross section at the two energies.

\begin{figure}
\resizebox{0.5\textwidth}{!}{
  \includegraphics{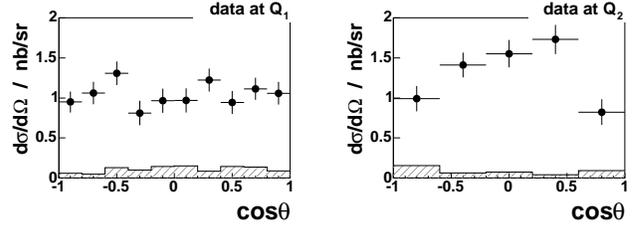}
} \caption{Experimental angular distributions of the \hefeta{}
reaction at the two excess energies. The error bars attached to
the points are statistical, while the systematic ones are drawn as
histograms.} \label{fig:final}
\end{figure}

Of the corresponding angular distributions depicted in
fig.~\ref{fig:final}, the one obtained at excess energy $Q_1$
appears isotropic, whereas that at~$Q_2$~reveals a strong angular
dependence. Due to the identical nature of the deuterons in the
initial state, the cross section must be symmetric in $\cos\theta$
and our results are consistent with this within the error bars.

\begin{figure}
\resizebox{0.5\textwidth}{!}{
  \includegraphics{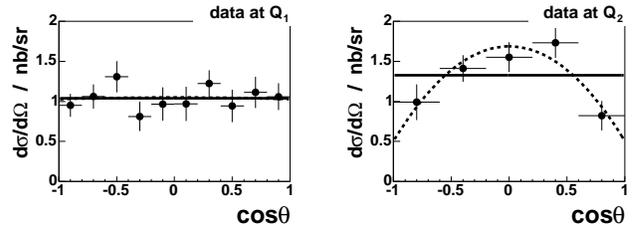}
} \caption{Fits of a constant and a second order polynomial to the
experimental \hefeta{} angular distributions where the
angular-dependent systematic errors have been added in quadrature.
At $Q_1$ there is no evidence for any non-isotropy whereas at
$Q_2$ the $\textit{Pol2}(\cos\theta)$ fit describes the data much
better.} \label{fig:final_fits}
\end{figure}

In order to quantify the results, both angular distributions were
each fitted with two functions: a constant
$\textit{Pol0}(\cos\theta) = C_0$ and a symmetric second order
polynomial $\textit{Pol2}(\cos\theta) = C_0\,\left(
1+C_2\cos^2\theta \right)$. For fitting purposes, statistical and
angular-dependent systematic errors were added in quadrature. The
resulting fitted curves are superimposed on the experimental data
in fig.~\ref{fig:final_fits} while the values of the parameters
are collected in table~\ref{tab:fitresults}. From this procedure
it is clear that at the higher energy there is a significant
negative value for $C_2$.

\begin{table}
  \caption{Results of constant and quadratic fits to
  the experimental angular distributions.}
  \label{tab:fitresults}
\begin{tabular}{ccc}
& \multicolumn{2}{c}{ fit of $\textit{Pol0}(\cos\theta)$} \\
\hline\noalign{\smallskip}
& $C_0$ & $\chi^2/ndf$  \\
\hline\noalign{\smallskip}
$Q_1$ & $1.04\pm 0.06$ & $\phantom{1}5.6/9=0.63$ \\
$Q_2$ & $1.33\pm 0.09$ & $16.1/4=4.03$ \\
\hline\noalign{\smallskip}
&&\\
\end{tabular}
\begin{tabular}{cccc}
& \multicolumn{2}{c}{fit of $\textit{Pol2}(\cos\theta)$} \\
\hline\noalign{\smallskip}
& $C_0$ & $C_2$ &  $\chi^2/ndf$ \\
\hline\noalign{\smallskip}
$Q_1$ & $1.05\pm 0.10$ & $-0.02\pm0.18$ & $5.6/8=0.70$\\
$Q_2$ & $1.69\pm 0.13$ & $-0.70\pm0.16$ & $2.7/3=0.91$\\
\hline\noalign{\smallskip}
\end{tabular}
\end{table}

\section{Comparison to World data}
\label{sec:compar} \setcounter{equation}{0}

The results of this work should be compared with the existing
World data on the \hefeta{} total cross sections. However, whereas
the near-threshold data of ref.~\cite{Frascaria94} were taken with
an unpolarised beam, the values quoted in ref.~\cite{Willis97}
were obtained with a polarised deuteron beam of helicity $m=\pm1$.
At threshold, where the $\eta\,\hef$ system is in the $s$-wave,
this gives the only non-vanishing cross section. The group indeed
observed no signal of \hefeta{} for a beam polarisation $m=0$ at
$p_{\h}=48$~MeV/c, a region where we also find an isotropic
differential cross section. Unfortunately there is no record in
the publication of such an absence at higher
energies~\cite{Willis97}. The evaluation of the unpolarised total
cross section from the polarised data depends on the partial waves
composition assumed and cannot be performed in a completely
model-independent way, as long as no additional polarised data is
available.

\subsection{Comparison of total cross section}
\label{4_1}

In the analysis of the SPESIII experiment~\cite{Willis97}, it was
assumed that only $s$-waves were significant. In this case the
unpolarised total cross section was taken to be $\frac{2}{3}$ of
the $m=\pm 1$ cross section that was actually measured in the
experiment. However, the angular distribution shown in
fig.~\ref{fig:final} is clear evidence of the influence of higher
partial waves at the excess energy $Q_2$. Now it is shown in the
appendix that, to order $p_{\eta}^2$, there are only three partial
wave amplitudes that contribute to unpolarised cross section and
the alignment. These correspond to final $\eta$ $s$--waves
($A_0$), $p$--waves ($C$), and $d$-- waves ($A_2$). The anisotropy
could be due to $s$--$d$ interference or be a pure $p$--wave
effect and these two possibilities give rise to a different
relation between the $m=\pm1$ and the unpolarised cross section.
This can be seen from the expressions in eq.~(\ref{eq:cterm}),
derived in the appendix, where we have chosen a convenient
normalisation for which
\begin{eqnarray}
\nonumber
  \left(\frac{\dd\sigma}{\dd\Omega}\right)_{\!\!m=\pm 1} &=&
\frac{p_{\h}}{p}\left[|A_0|^2+
2\mathrm{Re}(A_0^*A_2)\,p_{\h}^2\,P_2(\cos\theta)\right.\\
\nonumber &&\hspace{2.3cm}+\left.
  \half|C|^2p_{\h}^2\sin^2\theta + {\cal O}(p_{\h}^3)\right],\\
  \left(\frac{\dd\sigma}{\dd\Omega}\right)_{\!\!m=0\:\:\:} &=&
\frac{p_{\h}}{p}\left[|C|^2p_{\h}^2\sin^2\theta  + {\cal
O}(p_{\h}^3) \right]\:.
  \label{eq:cterm}
\end{eqnarray}

\begin{figure}[!htb]
  \resizebox{0.5\textwidth}{!}{
    \includegraphics{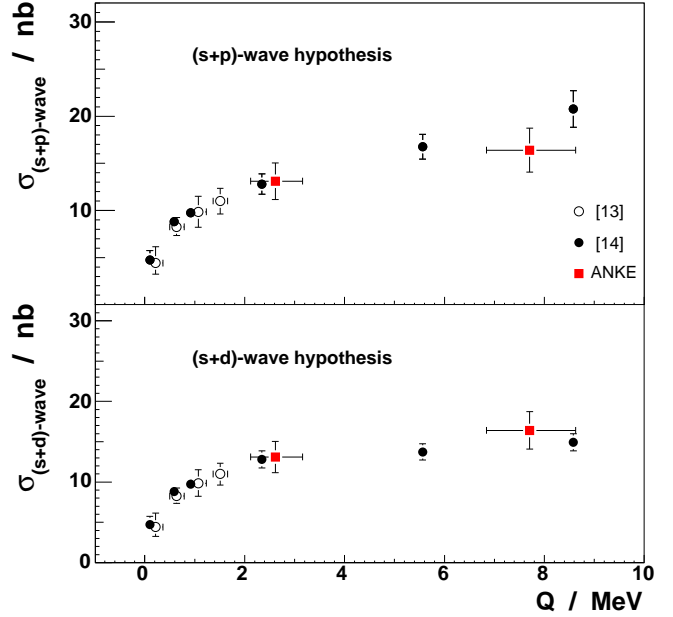}
  }
\caption{World data on the unpolarised \hefeta{} total cross
section. Our results (squares) are compared to the measurements of
refs.~\cite{Frascaria94} and \cite{Willis97}, the latter being
corrected for the unmeasured $m=0$ cross section, as described in
the text. In the upper panel, uncertainties in the correction
procedure leads to larger error bars being assigned to the data of
ref.~\cite{Willis97}.}
  \label{fig:sigmaTotal}
\end{figure}

We have fit our data, to the order we keep, first by assuming the
$s+p$ scenario ($A_2=0$) and determining the $|C|^2$ factor. This
factor was then applied to calculate  $\sigma(m=0)$ at the momenta
of two last SPESIII points~\cite{Willis97}. These unmeasured
contributions were then added to the two SPESIII points to
calculate the full unpolarised cross section. For the $s+d$
hypothesis ($C=0$) there is no such contribution and
$\sigma(\textrm{unpolarised})=\fmn{2}{3}\sigma(m=\pm1)$, as
calculated in the original paper. Fig.~\ref{fig:sigmaTotal}
presents the World data set and the difference between the two
approaches to the SATURNE data represents the systematic
uncertainty due to the unmeasured $m=0$ cross section.

It is seen from the figure that our results are, within errors,
consistent with the SATURNE data independent of the assumption
that we make regarding the partial wave composition.

\subsection{Consequences for partial amplitudes}

The rapid variation of the average production amplitude with
momentum close to threshold is the signal for a strong
$\eta\,\alpha$ final state interaction. It is common in such a
case to parameterise the $s$--wave amplitude in terms of the
complex scattering length $a_{\eta\alpha}$~\cite{Goldberger64};
\begin{equation}
f_s = \frac{f_B}{1-ip_{\h}a_{\eta\alpha}}\:, \label{scatlength}
\end{equation}
where $f_B$ is assumed to change little with momentum.

\begin{figure}
  \resizebox{0.5\textwidth}{!}{
    \includegraphics{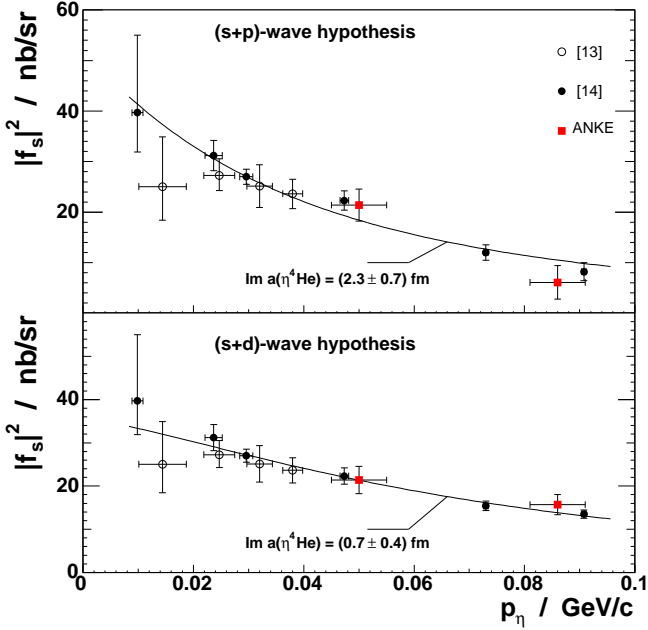}% created with macro util.C
  }
\caption{The square of the magnitude of the \hefeta{} $s$--wave
amplitude calculated with different assumptions concerning the
partial wave decomposition, as discussed in the text. Upper panel:
only $s$-- and $p$--wave contributions. Lower panel: $s$-- and
$d$--wave contributions. The curves represent fits using the
$s$--wave scattering length formula of eq.~(\ref{scatlength}).}
  \label{fig:ampl_fig2}
\end{figure}

To use the scattering length \emph{ansatz} one has first to
project out the contribution from the final $s$--wave ($|A_0|^2$
in eq.~(\ref{eq:cterm})). Now the angular distribution alone does
not provide sufficient information to do this. One can, as
discussed in sect.4.1 and shown in fig.~\ref{fig:sigmaTotal},
consider special cases where one of the terms vanish to see what
effect it would have on the $s$--wave cross section and hence on
the extraction of the scattering length from the data. The results
of this investigation of the $s+p$--wave and $s+d$--wave
hypotheses are displayed in fig.~\ref{fig:ampl_fig2}.

It is not possible from a single distribution such as that in
fig.~\ref{fig:ampl_fig2} to determine two real parameters; the
error bars become strongly correlated. The authors of
ref.~\cite{Willis97} did a combined optical model fit to all the
near--threshold \heteta{} and \hefeta{} data, assuming that only
$s$--wave production occurs, and this resulted in a value
$a_{\eta\alpha}=(-2.2+1.1i)\:$fm. In this analysis the real part
of $a_{\eta\alpha}$ was fixed mainly by the $\eta\,^3\mathrm{He}$
data and so the scattering length fits of fig.~\ref{fig:ampl_fig2}
have been carried out by varying just
$\textrm{Im}(a_{\eta\alpha})$. This resulted in $a_{\eta\alpha}=
(-2.2 + 2.3i)\:$fm and $(-2.2 + 0.7i)\:$fm respectively for the
$s+p$ and $s+d$ assumptions. One can only attempt to separate
these solutions by having measurements with a polarised beam,
where eq.~(\ref{A8}) allows $|C|^2$ to be extracted directly,
\emph{e.g.}, from $t_{22}$.

\section{Conclusions}
\label{sec:concl} \setcounter{equation}{0}

By measuring the angular distribution of the \hefeta{} reaction at
two different energies, we have shown that higher partial waves
affect the differential cross section at lower values of excess
energy than for the corresponding \heteta{} reaction. The total
cross sections link well with the results of previous
measurements, though there is some ambiguity at the higher
energies where the earlier data were taken with a polarised
deuteron beam of helicity $m=\pm 1$. The uncertainty in the origin
of the higher partial waves is reflected in the estimation of the
$\eta\,\alpha$ scattering length from the data. Fixing the
position of the pole in the $\eta\,\alpha$ scattering amplitude,
which determines whether or not there is a quasi-bound state,
requires an even more extensive data set, including especially
more measurements with a polarised deuteron beam~\cite{mariola},
as well as data from below the $\eta \alpha$ threshold analogous
to the studies for the $\eta ^3$He system described in
ref.~\cite{Pfeiffer}.

The only theoretical model designed to describe the \hefeta{}
reaction in a two--step approach has only been evaluated in the
$s$-wave limit~\cite{Faldt96}. It would be helpful to extend such
calculations to higher waves to provide an indication as to which
of these first becomes significant.

\begin{acknowledgement}
Correspondence with Nicole Willis on the data of
ref.~\cite{Willis97} is gratefully acknowledged, as is the help of
J.~Smyrski in setting up the MWDCs. This work was carried out
within the framework of the ANKE collaboration~\cite{AC} and
supported by the Forschungszentrum J{\"u}lich and the European
Community -- Access to Research Infrastructure action of the
Improving Human Potential Programme.%
\end{acknowledgement}
%
%%%%%%%%%%%%%%%%%%%%%%%%%%%%%%%%%%%%%%%%%%%%%%%%%%%%%%%%%%%%%%%%%%
%
\begin{appendix}
\section{\hspace{-1.1mm}ppendix: The \bmath{dd\to\alpha\eta} amplitudes}
 \label{appendix}
\setcounter{equation}{0}

In order to be able to isolate the $s$--wave amplitude from data
on the production of pseudoscalar mesons in reactions such as
$dd\to\alpha\eta$ or $dd\to\alpha\pi^0$, one has to make
measurements of deuteron analysing powers as well as the
differential cross section. The resulting analysis requires an
understanding of the relationship between the amplitude structure
and the observables, which is summarised in this appendix.

Due to the identical nature of the incident deuterons, only three
independent scalar amplitudes are necessary to describe the spin
dependence of the reaction. If we let the incident deuteron cms
momentum be \bmath{p} and that of the $\eta$ be \bmath{p_{\eta}},
then one choice for the structure of the transition matrix
$\mathcal{M}$ is%
\footnote{ Parity conservation together with Bose symmetry
prohibits the appearance of a term
$({\bmath{\epsilon}}_1\cdot{\bmath{\epsilon}}_2)$ and a structure
such as $({\bmath{\epsilon}}_1\cdot{\hat{\bmath{p}}})
{\bmath{\epsilon}}_2\!\cdot\!({\bmath{p}_{\eta}}\times\hat{\bmath{p}})
-({\bmath{\epsilon}}_2\cdot{\hat{\bmath{p}}})
{\bmath{\epsilon}}_1\!\cdot\!({\bmath{p}_{\eta}}\times\hat{\bmath{p}})$
can be rewritten as a linear combination of the first two terms in
Eq. (\ref{A1}) (\emph{c.f.} eq.~(B.7) in ref.~\cite{chreport}).}:
\begin{eqnarray}
\nonumber M&=&A({\bmath{\epsilon}}_1\times{\bmath{\epsilon}}_2)
\cdot\hat{\bmath{p}}
+B({\bmath{\epsilon}}_1\times{\bmath{\epsilon}}_2)\cdot
\left[\hat{\bmath{p}}\times({\bmath{p}_{\eta}}\times\hat{\bmath{p}})\right]
({\bmath{p_{\eta}}}\cdot{\hat{\bmath{p}}})\\
&&\hspace{-2mm}+C\left[({\bmath{\epsilon}}_1\cdot{\hat{\bmath{p}}})
{\bmath{\epsilon}}_2\!\cdot\!({\bmath{p}_{\eta}}\times\hat{\bmath{p}})
+({\bmath{\epsilon}}_2\cdot{\hat{\bmath{p}}})
{\bmath{\epsilon}}_1\!\cdot\!({\bmath{p}_{\eta}}\times\hat{\bmath{p}})
\right], \label{A1}
\end{eqnarray}
where the ${\bmath{\epsilon}}_i$ are the polarisation vectors of
the two deuterons. Note that $\mathcal{M}$, which is a
pseudoscalar due to the $\eta$ parity, is invariant under the
transformation ${\bmath{\epsilon}}_1 \rightleftharpoons
{\bmath{\epsilon}}_2$, ${\bmath{p}} \to -{\bmath{p}}$, as required
by Bose symmetry. The scalar amplitudes $A$, $B$, and $C$ are
functions of ${\bmath{p}_{\eta}}^2$, ${\bmath{p}}^2$, and
$({\bmath{p}_{\eta}}\cdot{\bmath{p}})^2=p_{\eta}^2p^2\cos^2\theta$,
where $\theta$ is production angle of the $\eta$ meson.

Following the usual convention of letting ${\bmath{p}}$ lie along
the $z$--direction and ${\bmath{p}_{\eta}}$ to be in the $x$--$z$
plane, the transition matrix reduces to:
\begin{eqnarray}
\nonumber
\mathcal{M}&=&A\left[\epsilon_{1x}\epsilon_{2y}-\epsilon_{1y}\epsilon_{2x}
\right]+ Bp_{\eta}^2\sin\theta\cos\theta
\left[\epsilon_{1y}\epsilon_{2z}-\epsilon_{1z}\epsilon_{2y}
\right]\\
&&- Cp_{\eta}\sin\theta
\left[\epsilon_{1z}\epsilon_{2y}+\epsilon_{1y}\epsilon_{2z}
\right].\label{A2}
\end{eqnarray}

If we assume that deuteron--2 is unpolarised, the remaining
polarisation information is contained within the density matrix:
\begin{equation}\label{A3}
\mathcal{Z}=\sum_{m_2}\mathcal{M}^{\dagger}\mathcal{M}\,.
\end{equation}

Using the explicit form of eq.~(\ref{A2}), and working in the
spherical basis, the unpolarised intensity ($I$) and the vector
($t_{1i}$) and tensor ($t_{2i}$) analysing powers are obtained by
taking the trace of $\mathcal{Z}$ with the unit matrix, and the
vector and tensor projection operators $\Omega_{1i}$ and
$\Omega_{2i}$~\cite{Hamilton}. This leads to the
expressions:
\begin{eqnarray}
\nonumber
I&=&\fmn{2}{3}\left(|A|^2+|B|^2p_{\eta}^4\sin^2\theta\cos^2\theta
+|C|^2p_{\eta}^2\sin^2\theta\right),\\
\nonumber
I\,t_{20}&=&\fmn{1}{3\sqrt{2}}\left(2|A|^2-|B|^2p_{\eta}^4\sin^2\theta\cos^2\theta\right.\\
\nonumber
&&\left.-|C|^2p_{\eta}^2\sin^2\theta-6\,\textrm{Re}\{B^*C\}p_{\eta}^3\sin^2\theta\cos\theta\right),\\
\nonumber
I\,t_{21}&=&-\fmn{1}{\sqrt{3}}\,\textrm{Re}\left\{A^*(Bp_{\eta}\cos\theta+C)\right\}p_{\eta}\sin\theta,\\
\nonumber I\,t_{22}&=&\fmn{1}{2\sqrt{3}}\,|Bp_{\eta}\cos\theta-C|^2p_{\eta}^2\sin^2\theta,\\
\nonumber
I\,it_{11}&=&\fmn{1}{\sqrt{3}}\,\textrm{Im}\left\{A^*(Bp_{\eta}\cos\theta+C)\right\}p_{\eta}\sin\theta,\\
I\,t_{10}&=&0\,.\label{A4}
\end{eqnarray}

The term proportional to $C$ is the only one that changes sign
under $\vec{p_{\h}}\to-\vec{p_{\h}}$ and thus represents odd
$\eta\,\alpha$ partial waves; the other amplitudes correspond to
even waves. Of these, the sole term that survives at threshold,
and which therefore contains the $\eta\,\alpha$ $s$--wave, is
proportional to $A$ so that the tensor analysing power
$t_{20}=+1/\sqrt{2}$ at threshold and inspection of eq.~(\ref{A4})
shows this to be true more generally at $\theta=0$.

For simplicity of presentation, we have adopted a notation in
eq.~(\ref{eq:cterm}) whereby the unpolarised differential cross
section is related to the amplitudes by
\begin{equation}
\frac{\dd\sigma}{\dd\Omega}=\frac{p_{\eta}}{p}\,I\:. \label{A5}
\end{equation}

The azimuthally symmetric differential cross sections for defined
initial polarisation used in eq.~(\ref{eq:cterm}) depend only then
on $t_{20}$ and the unpolarised cross section through
\begin{eqnarray}
\nonumber
  \left(\frac{\dd\sigma}{\dd\Omega}\right)_{\!\!m=\pm 1} &=&
\left(1+t_{20}/\sqrt{2}\right)\,\left(\frac{\dd\sigma}{\dd\Omega}\right),\\
  \left(\frac{\dd\sigma}{\dd\Omega}\right)_{\!\!m=\pm 0} &=&
\left(1-\sqrt{2}\,t_{20}\right)\,\left(\frac{\dd\sigma}{\dd\Omega}\right).
\label{A6}
\end{eqnarray}

Since for our experimental data we are interested in describing
the first deviations from an $s$-wave behaviour, we shall only
keep terms that contribute to the observables up to order
$p_{\h}^2$. Though to this order $B$ and $C$ can be taken as
constant at their threshold values, there can be an angular
dependence in the $A$ amplitude which, to this order, may be
written as the truncated Legendre expansion:
\begin{equation}
\label{A7} A= A_0+A_2\,p_{\h}^2\,P_2(\cos\theta)\:.
\end{equation}

To order $p_{\eta}^2$ therefore,
\begin{eqnarray}
\nonumber
I&=&\fmn{2}{3}\left(|A_0|^2+2\,p_{\eta}^2\textrm{Re}\{A_0^*A_2\}\,P_2(\cos\theta)\right.\\
\nonumber&&\hspace{4cm}\left.+|C|^2p_{\eta}^2\sin^2\theta\right),\\
\nonumber
I\,t_{20}&=&\fmn{1}{3\sqrt{2}}\left(2|A_0|^2+4\,p_{\eta}^2
\textrm{Re}\{A_0^*A_2\}\,P_2(\cos\theta)\right.\\ \nonumber
&&\hspace{4cm}\left.-|C|^2p_{\eta}^2\sin^2\theta\right),\\
\nonumber
I\,t_{21}&=&-\fmn{1}{\sqrt{3}}\,p_{\eta}\sin\theta\,\left[\textrm{Re}\{A_0^*B\}p_{\eta}\cos\theta
+\textrm{Re}\{A_0^*C\}\right],\\
\nonumber I\,t_{22}&=&\fmn{1}{2\sqrt{3}}\,|C|^2p_{\eta}^2\sin^2\theta,\\
\nonumber
I\,it_{11}&=&\fmn{1}{\sqrt{3}}\,p_{\eta}\sin\theta\,\left[\textrm{Im}\{A_0^*B\}p_{\eta}\cos\theta
+\textrm{Im}\{A_0^*C\}\right],\\
I\,t_{10}&=&0\,,\label{A8}
\end{eqnarray}
where the $s$--wave amplitude $A_0$ will have an additional $p_{\eta}$
dependence arising from the $\eta$-$\alpha$ final state interaction.

Thus measurements of the angular distributions of the unpolarised
cross section, $t_{20}$, and $it_{11}$, would allow one to extract
the values of $|A_0|^2$, Re$(A_0^*A_2)$, $|C|^2$, Im$(A_0^*B)$,
and Im$(A_0^*C)$. This would then lead to two two--fold
%\newpage\noindent%
ambiguities that could only be resolved by the measurement of
$t_{21}$. To this order in the momentum expansion, the $t_{22}$
information is not independent, though it would provide some check
on the systematics arising for example from the background
subtraction and/or on the convergence of the momentum expansion.
\end{appendix}

%
%%%%%%%%%%%%%%%%%%%%%%%%%%%%%%%%%%%%%%%%%%%%%%%%%%%%%%%%%%%%%%%%%%
%
%\newpage

\end{document}